\documentclass[epsfig,preprint2]{aastex}

\shorttitle{The central binary and surrounding nebulae of V1016 Cyg}
\shortauthors{Brocksopp et al.}

\begin{document}

\title {The central binary and surrounding nebula of the symbiotic star V1016 Cygni}
\author{C. Brocksopp, M.F. Bode}
\affil{Astrophysics Research Institute, Liverpool John Moores University, Twelve Quays House, Egerton Wharf, Birkenhead CH41 1LD, UK}
\email{cb@astro.livjm.ac.uk, mfb@astro.livjm.ac.uk}
\author{S.P.S. Eyres}
\affil{Centre for Astrophysics, University of Central Lancashire, Preston PR1 2HE}
\email{spseyres@uclan.ac.uk}
\author{M.M. Crocker}
\affil{Jodrell Bank Observatory, University of Manchester, Macclesfield, Cheshire SK11 9DL}
\email{mc@jb.man.ac.uk}
\author{A.R. Taylor}
\affil{Dept. of Physics and Astronomy, University of Calgary, 2500 University Dr. N.W., Calgary, Alberta T2N 1N4, Canada}
\email{russ@ras.ucalgary.ca}

\begin{abstract}
We present {\it HST}/WFPC2 images of the symbiotic star V1016 Cyg which, for the first time, directly explore the inner regions of the nebula to AU scales at optical wavelengths. They also suggest that the [O {\sc iii}] $\lambda\lambda$4959,5007 emission takes place in a bipolar outflow. We use these images to determine the position of the two stars and hence a projected binary separation of $84\pm2$ AU (assuming a distance of 2 kpc) and a position angle of $143.5\pm10^{\circ}$. Furthermore, we combine our images with VLA radio imaging which enables diagnostic tests to be applied and properties of the circumstellar nebula to be determined. Finally we have obtained archive {\it HST}/STIS spectra of V1016 Cyg with which we were able to spatially resolve the source at various positions in the nebula. This enabled discovery of the ultraviolet counterpart to the $\sim$25 arcsec extended emission found by Bang et al. (1992).

\end{abstract}

\keywords{binaries: symbiotic --- stars: individual (V1016 Cyg) --- circumstellar matter --- radio continuum: stars}

\section{Introduction}
V1016 Cyg is a symbiotic star which falls into the D(usty)-type sub-category, containing a Mira variable as its late-type cool component and a white dwarf as the hot component (e.g. Taranova \& Yudin 1983). There are a number of distance estimates in the literature, ranging from 2.1 to 10 kpc (Watson et al. 2000, and references therein). The binary period of the system has been suggested to be 80$\pm$25 years from spectropolarimetry (Schild \& Schmid 1996) or more recently as $\sim15$ years from optical photometry (Parimucha et al. 2000). Infrared photometry also shows a $\sim$450 day periodicity which has been attributed to the pulsations of the Mira (Harvey 1974).

\begin{table*}
\center
\caption{A list of the filters, their properties and the exposure times used in our {\it HST}/WFPC2 observations. $\dagger$ refers to those images which were obtained in the dithered mode in order to improve the spatial resolution.}
\label{filters}
\vspace{0.5cm}
\begin{tabular}{lccccc}
\hline
Filter&Exposure&$\lambda$&$\Delta\lambda$& Peak $\lambda$&Scientific features\\
&time (s)&(\AA)&(\AA)&(\AA)&and wavelengths (\AA)\\
\hline
F218W$^\dagger$&40&2136&355.9&2091&Hot continuum\\
F218W&100&2136&355.9&2091&Hot continuum\\
F437N$^\dagger$&40&4369&25.2&4368&[O {\sc iii}] $\lambda$4364\\
F437N&510&4369&25.2&4368&[O {\sc iii}] $\lambda$4364\\
F469N$^\dagger$&20&4695&24.9&4699&He {\sc ii} $\lambda$4686\\
F469N&100&4695&24.9&4699&He {\sc ii} $\lambda$4686\\
F502N$^\dagger$&20&5012&26.8&5009&[O {\sc iii}] $\lambda\lambda$4959,5007\\
F502N&100&5012&26.8&5009&[O {\sc iii}] $\lambda\lambda$4959,5007\\
F547M$^\dagger$&20&5454&486.6&5362&Cool continuum\\
F656N$^\dagger$&10&6562&22.0&6561&H$\alpha$ $\lambda$6563\\
\hline
\end{tabular}
\end{table*}

The system is well known for its nova-like optical outburst in 1965 and has since been dubbed a symbiotic nova (see e.g. Parimucha et al. 2000 and references therein). The outburst is thought to be the result of a thermonuclear runaway on the surface of the white dwarf as it accretes material emitted by the Mira (Miko$\mbox{\l}$ajewska \& Kenyon 1992). As with classical novae, symbiotic novae may be accompanied by a radio outburst -- while radio monitoring did not take place in 1965, V1016 Cyg is certainly one of the brightest radio sources among the symbiotic systems (e.g. Belczy\'nski et al. 2000).

The source has been imaged previously in the radio and emits a spectrum typical of thermal bremsstrahlung. While originally thought to be produced in an expanding shell as a result of the 1965 outburst, Watson et al. (2000) used MERLIN observations to show that it is more likely that the peaks in the radio maps are the result of colliding winds between the two stellar components. Despite a number of similarities between V1016 Cyg and another D-type symbiotic, HM Sagittae, V1016 Cyg does not show any evidence for non-thermal emission which has been detected in the former system (Richards et al. 1999).

\begin{table}
\center
\caption{Pixel ranges used in the images shown in Figs.~\ref{fig1} and \ref{fig2}. Flux units for all filters are erg/s/cm$^2$/\AA/arcsec$^2$; the F502N/F437N ratio map is dimensionless.}
\vspace{.5cm}
\label{greyscale}
\begin{tabular}{lcc}
\hline
Filter&$\sigma$& Max. Flux\\
\hline
F218W&$5.3\times10^{-14}$&$4.5\times10^{-13}$\\
F437N&$3.6\times10^{-15}$&$4.8\times10^{-14}$\\
F469N&$7.2\times10^{-15}$&$1.2\times10^{-13}$\\
F502N&$2.5\times10^{-15}$&$5.6\times10^{-14}$\\
F547M&$1.5\times10^{-16}$&$2.7\times10^{-15}$\\
F656N&$1.2\times10^{-15}$&$2.1\times10^{-14}$\\
Ratio&0 (min.) &$287$\\
\hline
\end{tabular}
\end{table}

\begin{figure*}
\begin{center}
\epsscale{2}
%\vspace*{-2.5cm}
\plotone{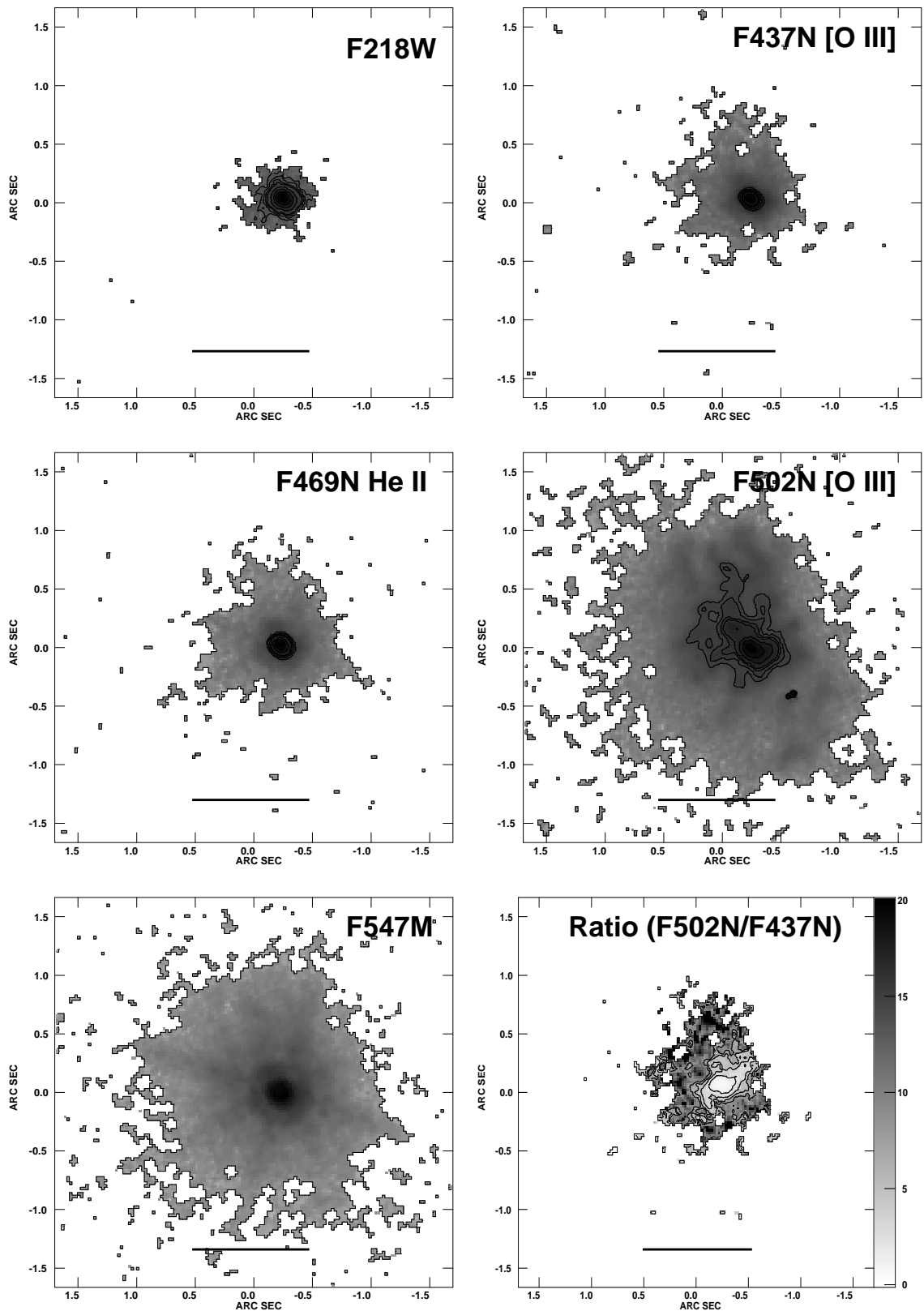}  
\caption{{\it HST} images of V1016 Cyg in the F218W (`hot continuum', top left), F469N (He {\sc ii} $\lambda 4686$, middle left), F547M (`cool continuum', bottom left), F437N ($\lambda 4364$, top right) F502N ($\lambda\lambda 4959, 5007$, middle right) filters. The sixth plot (bottom right) is the ratio of filters F502N:F437N. Greyscale pixel ranges for these images are listed in Table~\ref{greyscale}. The same axes have been used for each of the plots to aid comparison; the black line across each plot corresponds to 2000 AU, assuming a distance of 2kpc to V1016 Cyg. Contours represent 3$\sigma \times$ 1, 2, 4, 8, 16, 32 in the images (also 3$\sigma \times$ 0.667, 1.5 in F502N. See text for further details.}
\label{fig1}
\end{center}
\end{figure*}

\begin{figure*}
\begin{center}
%\vspace*{-2.5cm}
\includegraphics[angle=90,height=7cm]{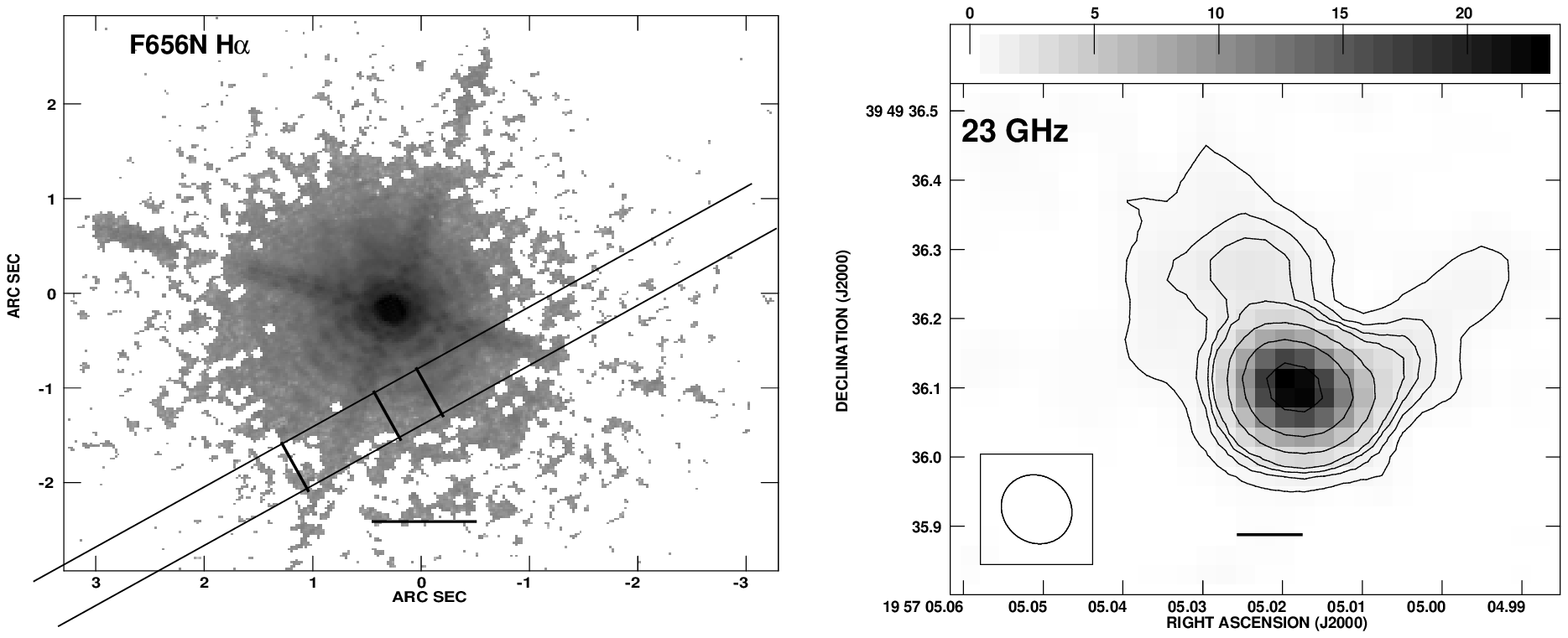}
\caption{{\it HST} H$\alpha$ image taken with the F656N filter ($\lambda 6563$, left) and the 23 GHz radio map (right). The greyscale pixel range for the H$\alpha$ image is listed in Table~\ref{greyscale}; the approximate position of the slit is indicated (the position angle of which is coincidentally aligned roughly with that of the binary), as well as the extraction positions of the spectra shown in Fig.~\ref{spectra} (see Section 3.5). Contours on the radio map indicate radio flux density values of 3, 6, 12, 24, 48 and 96 $\times \sigma$ where $\sigma=0.19$ mJy. The black horizontal lines correspond to 2000 AU and 200 AU in the H$\alpha$ image and radio map respectively, assuming a distance of 2kpc to V1016 Cyg.}
\label{fig2}
\end{center}
\end{figure*}

The majority of optical papers in the literature have concentrated on spectroscopic observations and there is little in the way of imaging. What there is has revealed an extended nebula (Bang et al. 1992, Corradi et al. 1999) but no counterpart in either the infrared or ultraviolet has been observed. The resolution of ground-based optical telescopes has been insufficient for any observations of the inner nebula at sub-arcsecond scales to have taken place.

In this paper we present {\it HST}/WFPC2 images of V1016 Cyg with resolution of $<$0.1 arcsec. We combine our results with VLA radio imaging and archival ultraviolet {\it HST}/STIS observations which enable diagnostic tests to be applied in order to determine various physical parameters.

\section{Observations}

\subsection{{\it HST}/WFPC2}
Our {\it HST} observations took place on 1999 November 4 as part of GO programme 8330 on symbiotic stars. Over the course of two orbits images were made in six different filters -- F218W, F437N, F469N, F502N, F547M, F656N -- details of which can be found in Table~\ref{filters} and in Biretta et al. (1996). These filters were chosen in order to probe the hot and cool continuum components of the symbiotic system (F218W and F547M respectively, see Eyres et al. 2001 for assumptions and justification), as well as various emission lines from which conditions within the nebula could be determined.

\begin{figure}
\begin{center}
\includegraphics[angle=90,height=6cm]{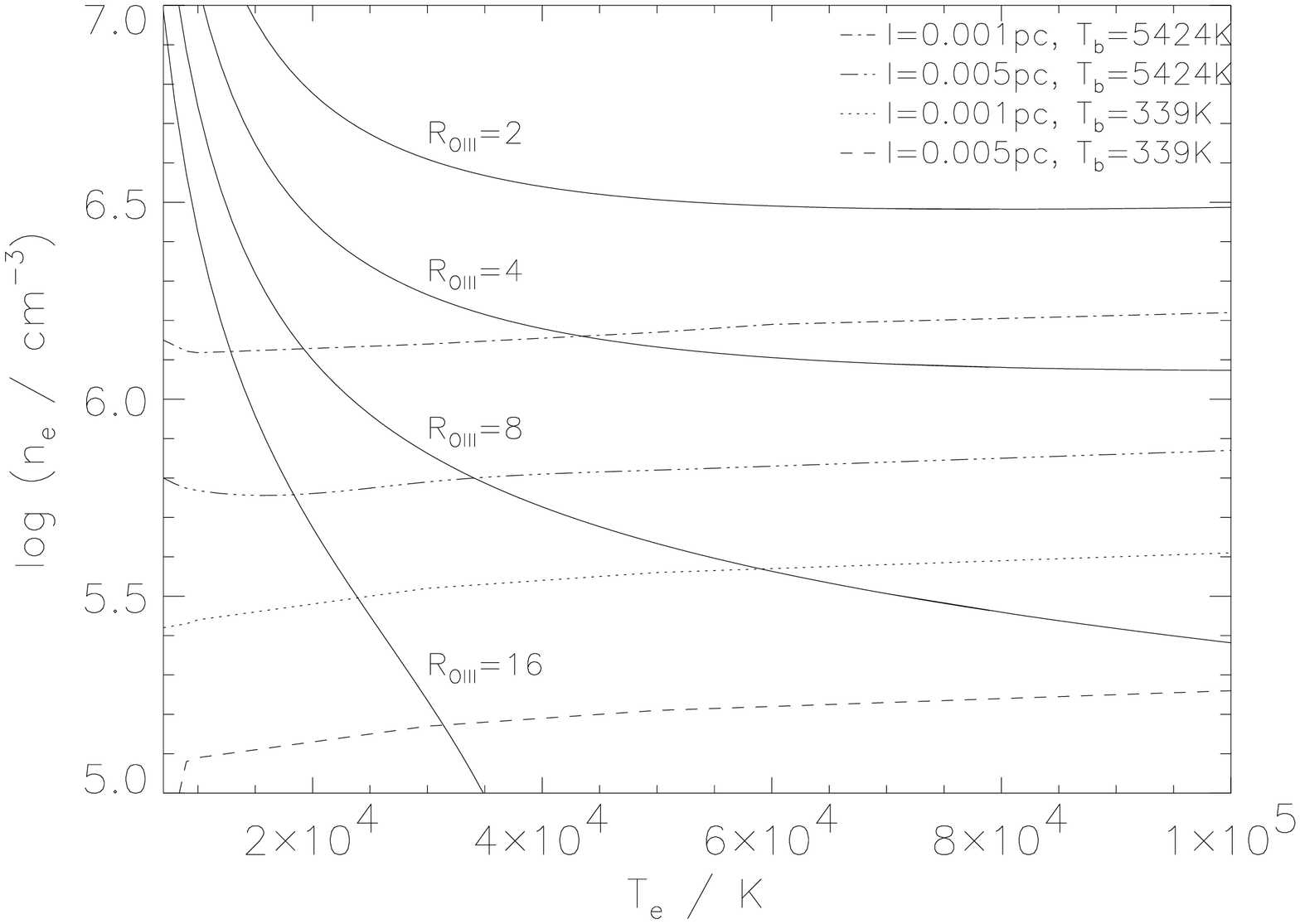}
\caption{Plots of $\log n_e$ against $T_e$ in terms of $R_{OIII}$ (solid lines) and $T_b$, $l$ (dotted/dashed lines). The different values of $R_{OIII}$, $T_b$ and $l$ are indicated next to the curves and in the key (see text for details).}
\label{nebula}
\end{center}
\end{figure}

The calibrated images were obtained from the {\it HST} archive. Since there were two observations in each filter it was possible to perform cosmic ray removal. A second pair of shorter exposures was also taken in each filter; this time the second image was `dithered' in order to allow full recovery of the spatial resolution (Biretta et al. 1996). The {\sc iraf} task {\sc drizzle} was used to recombine the images. The resultant pixel size of the dithered images was 0.02275 arcsec, compared with 0.0455 arcsec for the undithered images. We use the dithered images to obtain the results below. 

Each image has been de-reddened using the standard interstellar extinction curve of Whittet (1992) and published values of Rudy et al. (1990), who determined a value of E(B$-$V)=0.46$\pm$0.007 from the hydrogen lines. (See also Section 3.4.)

\subsection{{\it HST}/STIS}
We have also obtained two STIS spectra from the public archive (Proposal 8188, Cycle 8, P.I. C.D. Keyes). V1016 Cyg was observed on 2000 March 24 using the gratings G140L and G230L; the wavelength coverage is 1145--1740 and 1570--3180 $\mbox{\AA}$ with a spectral resolution of 0.60 and 1.58 \AA/pixel respectively. The slit width and length were 0.5 and 52 arcsec respectively with a pixel scale of 0.025 arcsec/pixel on the CCD chip.

\subsection{VLA}
The VLA observations were made on 1999 September 26 at 23 GHz. Whilst not strictly simultaneous with the WFPC2 observations we note that a $\sim40$ day period is short, relative to the likely timescales of variability in the source, and so therefore it is possible to compare them directly.

The images were reduced using standard flagging, calibration and imaging techniques within {\sc aips}. The primary calibrator was 3C286 (1331+305) which led to determination of the flux of the secondary calibrator, 2007+404, as $1.57\pm 0.07$ Jy. The complex gain solutions obtained were then applied to the target, V1016 Cyg for which a peak flux of $23.28\pm 0.17$ mJy/beam was measured. The final image was naturally weighted in order to improve the somewhat high noise levels at this frequency.

\section{Results and discussion}
\subsection{Images}

The resulting dereddened images can be seen in Figures~\ref{fig1} and \ref{fig2}. Fig.~\ref{fig1} shows the {\it HST} images in the F218W, F469N (He {\sc ii} $\lambda 4686$), F547M, F437N ([O{\sc iii}] $\lambda 4364$) and F502N ([O{\sc iii}] $\lambda\lambda 4959, 5007$), as well as the ratio of the two [O{\sc iii}] images. Fig.~\ref{fig2} shows the image taken with the H$\alpha$ filter, F656N and the 23 GHz radio map. North is to the top of the image in each case and an AU scale is given for each image at an assumed distance of 2kpc (the black line represents 2000 AU in each HST image and 200 AU in the radio and map).

The majority of the {\it HST} images show an approximately circular (spherical?) nebula ranging in size from 0.5--2 arcseconds in diameter; there are very few additional features. However in the case of F502N (Fig.~\ref{fig1}) the nebula is more asymmetric and, as we show in the next section, extended approximately perpendicular to the derived binary axis. We must consider the possibility that this is some form of bipolar outflow. Closer inspection of the brighter regions of the image reveals more intricate features with various filaments and enhanced `blobs'. Comparison with the larger-scale images of Corradi et al. (1999) is difficult due to the different resolution scales -- nonetheless, possible extension in the same direction as their work is suggested.

The 23 GHz radio map (Fig.~\ref{fig2}) shows a source with peak intensity $23.3\pm 0.17$ mJy/beam. Contours indicate flux density values of 3, 6, 12, 24, 48 and 96 $\times \sigma$ where $\sigma=0.19$ mJy. Structure is observed on a smaller scale compared with the optical; we note that the spatial resolution of the VLA and dithered {\it HST} images are comparable. Again an extension approximately perpendicular to the binary position angle (see below) is apparent, as well as an additional extension rotated $\sim 90^{\circ}$ clockwise from this.

The ratio map and the annotations on the H$\alpha$ image are explained in Sections 3.3 and 3.5.

\subsection{Stellar positions}

By making use of the dithering technique in order to recover full spatial resolution, we have been able to measure the position of the two stars of the symbiotic system. We use the argument presented in Eyres et al. (2001) to show that it is reasonable to assume that the hot component dominates the emission peak in the F218W filter and that the cool component dominates in the F547M filter.

We have used the {\sc AIPS} task {\sc imfit} to fit a Gaussian to these two images -- using of course the images in which the pixel size was 0.02275 arcsec. We note that in each case the Gaussian provided a good ($\chi^2_{\nu}\sim 1$) fit to the radial profiles and was in agreement (within the errors) with comparison point spread functions produced by the {\sc tinytim} software. The peak position was determined relative to 19 57 05 +39 49 36 -- the shift in the F218W image was $\alpha=0.09860\pm0.00004$, $\delta=0.9130\pm0.00005$; in the F547M image it was $\alpha=0.10028\pm0.00004$, $\delta=0.8789\pm0.00005$. Thus the hot component and the cool component are separated by 42.4 mas (to within an accuracy of $<1$ mas) at a position angle of 143.5$^{\circ}$ (to within an accuracy of $<10^{\circ}$).

This is only the second time the separation of the two stars in a symbiotic binary has been measured directly (HM Sge being the first -- Eyres et al. 2001) and such a measurement can place important constraints on estimates of the distance to and orbital period of the system. We used Kepler's third law to calculate that for a projected separation of 42 mas, an assumed distance of 2 kpc and an assumed total mass of 2M$_{\odot}$ the minimum orbital period is $\sim 544$ years. Since the system is non-eclipsing, thus ruling out a high inclination angle and increasing the separation, the period will actually be higher than this; obviously using one of the higher published distance estimates increases the period further still. In order to decrease the orbital period to $\sim 80$ years (Schild \& Schmid 1996) it is necessary to decrease the distance to $\sim 580$ pc. We note that even a period of 80 years is high compared with other symbiotic systems (Belczy\'nski et al. 2000) and suggest that {\em either} the distance to V1016 Cyg needs revising {\em or} the period of V1016 Cyg is unusually high {\em or} the published orbital periods for other symbiotic systems may be too low. However, we also note that these published values tend to be for S-type symbiotics which are expected to have shorter obital periods; D-type symbiotics are indeed expected to have periods on timescales of hundreds of years and so our result is consistent with this. A final possibility is that our {\it HST} observations took place around apastron of a highly eccentric orbit.

At this point we must add a note of caution to these results. The values for angular separation and position angle of the two components in V1016 Cyg and HM Sge are very similar (V1016 Cyg -- 42.4 mas, 143.5$^{\circ}$; HM Sge -- 40 mas, 130$^{\circ}$), perhaps suggesting some, as yet unknown, instrumental effect. We note that the {\it HST} roll angle was similar for the two targets (252$^{\circ}$ vs. 259$^{\circ}$). However measurement of the angular separation of the stellar components in CH Cyg (Eyres et al. 2002) has also been attempted and none was found, to within the errors of the method. This was consistent with the fact that the hot component was eclipsed by the giant at the time of the {\it HST} imaging, as shown by photometric monitoring, and adds credibility to the method. Furthermore, the position angle for the components of HM Sge is confirmed using a second method (Schmid et al. 2000). Perhaps we should not be too surprised by our results for V1016 Cyg as the observed similarities between them (e.g. position and separation of radio peaks; Watson et al. 2000, Eyres et al. 2001) would lead us to expect similar parameters. However further investigation of any possible instrumental cause is probably warranted.

\subsection{Nebular diagnostics}

The ratio of the [O{\sc iii}] lines in the nebula ($R_{OIII}$) depends on both electron temperature $T_e$ (K) and density $n_e$ (cm$^{-3}$) (Osterbrock 1989) according to:

\begin{equation}
R_{OIII}=\frac{j_{\lambda 4959}+j_{\lambda 5007}}{j_{\lambda 4363}}
=\frac{7.73exp[3.29\times 10^4T_e^{-1}]}{1+4.5\times 10^{-4}(n_eT_e^{-0.5})}
\end{equation}

Furthermore, the radio brightness temperature $T_b$ in Kelvins is given by
\begin{equation}
T_b=T_e(1-e^{-\tau_{\nu}})
\end{equation}

where the optical depth $\tau_{\nu}$ at frequency $\nu$ (GHz) is
\begin{equation}
\tau_{\nu}\approx8.24\times 10^{-2}T_e^{-1.35}\nu^{-2.1}n^2_el
\end{equation}

where $l$ (pc) is the path length through the nebula. Thus the three relationships can be used to plot the ratio $R_{OIII}$ and the brightness temperature $T_b$ as functions of $T_e$ and $n_e$. 

The ratio of the two [O{\sc iii}] maps has been obtained and is shown in Fig.~\ref{fig1}. We note that the [O{\sc iii}] lines dominate the emission in the filter bands used for our observations (e.g. Andrillat, Ciatti \& Swings 1982). Contours indicate $R_{OIII}\sim$ 2, 4, 6, 8 and increase outwards from the centre of the nebula. In the outer regions values increase further up to $R_{OIII}\sim$ 15 and the occasional pixel has a value higher still, although these should be treated with caution. The higher value contours have been omitted from the plot for clarity. Therefore, by using the same assumptions and considerations as Eyres et al. (2001) we can use our ratio map to determine the conditions of the nebula at different path lengths.

Curves for $R_{OIII}=$2, 4, 8, 16 have been plotted in $T_e$:$\log n_e$ space and can be seen in Fig.~\ref{nebula}. The brightness temperature of the radio emission has also been determined for the second-highest ($48\sigma$, i.e. $T_b\sim 5424$ K) and lowest ($3\sigma$ i.e. $T_b\sim 339$ K) contours in the radio map. The 23 GHz radio image suggests that the angular size of the nebula is 0.1--0.5 arcsec, corresponding to 0.001--0.005 pc (or $\sim$ 200--1030 AU assuming a distance of 2 kpc); unfortunately it is not possible to probe the more extended regions of the nebula without lower frequency and/or resolution radio images). Thus a further four curves corresponding to these values of $T_b$ and $l$ have also been included in Fig.~\ref{nebula}.

Fig.~\ref{nebula} suggests that $n_e$ increases with increased brightness temperature and decreases with increased path length, as we would expect. The regions of lowest $R_{OIII}$ are found right at the centre of the source (see ratio map in Fig.~\ref{fig1}) and it is clear that the regions of the nebula probed by our radio observations correspond to $R_{OIII}\le 8$ and electron temperatures in excess of 40-50000 K -- suggesting some shock heating may begin to become important here. The extended nebula is cooler, with $R_{OIII}\ge 8$; however, radio imaging at lower frequencies is required to probe these regions. We note that our results are not inconsistent with those of Schmid \& Schild (1990) who determine a similar value of $R_{OIII}$ but a temperature of $\sim 30000$ K. Their observations were at considerably lower spatial resolution and thus included the cooler regions of the nebula; our high resolution observations probe only the innermost and potentially significantly hotter parts of the nebula. 

\subsection{Extinction}

We make use of the fact that the dereddened H$\beta$ flux can be derived from the radio flux via the equation (Pottasch 1983):

\begin{equation}
S_{\nu}=2.51\times10^8T_e^{0.53}\nu^{-0.1}F({\mbox{H}\beta})\,\,\, \mbox{Jy}
\end{equation}

Thus from our measured values of the H$\alpha$ flux we can extrapolate the value for the measured H$\beta$ flux (Osterbrock 1989) and determine an estimate for the extinction across the nebula from the ratio of the radio to H$\alpha$ images:

\begin{equation}
E(B-V)=\frac{1}{1.46}\log\bigg[\frac{S_{\nu}}{F_{obs}(H\alpha)}\bigg]-7.15
\end{equation}

We note that, due to the poor absolute positional accuracy of the $HST$ compared with the VLA, it was necessary to assume that the peak flux position of both the radio map and H$\alpha$ image were identical and to re-align the $HST$ image accordingly. We also assumed that the radio emission is purely thermal and optically thin -- these are reasonable assumptions as outlined in Eyres et al. (2001).

The resultant map showed extinction peaks either side of the radio peak as one might expect from a hot source surrounded by an extended dusty nebula. However the values of E(B$-$V) derived (ranging from 1.4$\pm$1.0 to 2.5$\pm$2.0) are high compared with previous estimates (e.g. Birriel et al. 2000), possibly due to self-absorption in H$\alpha$, and noisy due to the poor S/N in the radio image and the broad PSF of the H$\alpha$ peak. 

\subsection{STIS spectra}

\begin{figure*}
\begin{center}
\includegraphics[angle=90,height=12cm]{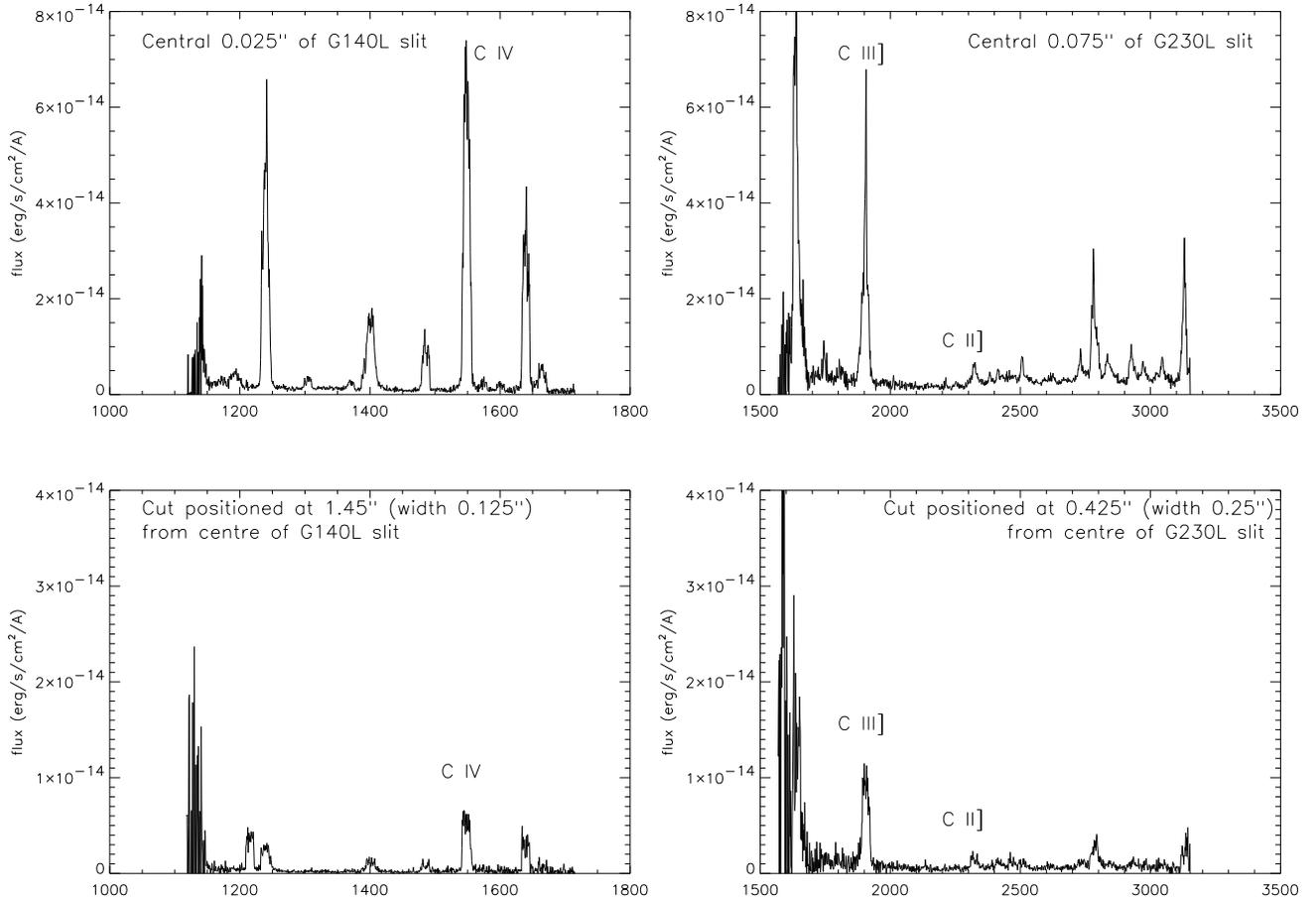}
\caption{Spectra from the centre of the G140L (top left) and G230L gratings (rows $y=$ 392 and 512 respectively). These correspond to the brighter regions of the nebula which were probed during the observations. The two bottom panels show sample spectra from further out in the nebula -- from rows $y=$ 450 and 495 for the G140L (bottom left) and G230L (bottom right) gratings respectively. These pixel values correspond to 1.45 arcsec (averaged over 0.125 arcsec) and 0.425 arcsec (averaged over 0.25 arcsec) from the slit centre respectively. The extraction positions are indicated by dashed lines on the H$\alpha$ image in Fig.~\ref{fig2}.}
\label{spectra}
\end{center}
\end{figure*}

The slit of the spectrograph lay at a position angle of 119.28$^{\circ}$ (from north measured through west) and had dimensions 0.5 x 52 arcsec; the approximate position of the slit is indicated in Fig.~\ref{fig2}. The zero position of the pointing was $\alpha=$ 19 57 05.0, $\delta=$39 49 36.5, corresponding to pixels $y=$ 392 and 512 for the G140L and G230L gratings respectively. Thus we were able to determine that the slit lay roughly parallel to the derived binary axis but that it lay across the nebula, not actually over either of the two stars.

Using the {\sc iraf} package {\sc x1d} we extracted one-dimensional spectra at many positions along the slit; the intervals were determined by the signal-to-noise possible at each position, with a greater degree of averaging taking place towards the outer parts of the slit, particularly for the G230L grating. Spectra centred on position $y=$ 392 and 512 (i.e. the slit centre) for the G140L and G230L gratings respectively) are shown in the top panels of Fig.~\ref{spectra} (averaged over 0.025 and 0.075 arcsec along the slit respectively); the bottom panels show sample spectra 1.45 arcsec and 0.425 arcsec from the slit centre respectively (averaged over 0.12 and 0.25 arcsec along the slit respectively). Each of these three positions is indicated by dotted lines on the H$\alpha$ image in Fig.~\ref{fig2}. We then used the {\sc splot} task to measure the integrated flux under various emission lines for each extracted spectrum and compared these values across the nebula. In order to do this direct comparison it was necessary to take into account the averaging of rows during the extraction. Thus each flux was divided by the number of rows averaged, the pixel scale and the slit width,  resulting in fluxes per square arcsecond.

%The He{\sc ii} $\lambda$1640 line was present in the spectra from both gratings and so could be used as a test -- while the line heights varied slightly, the integrated flux under the line remained approximately constant.

\begin{figure}
\begin{center}
\epsscale{1}
\plotone{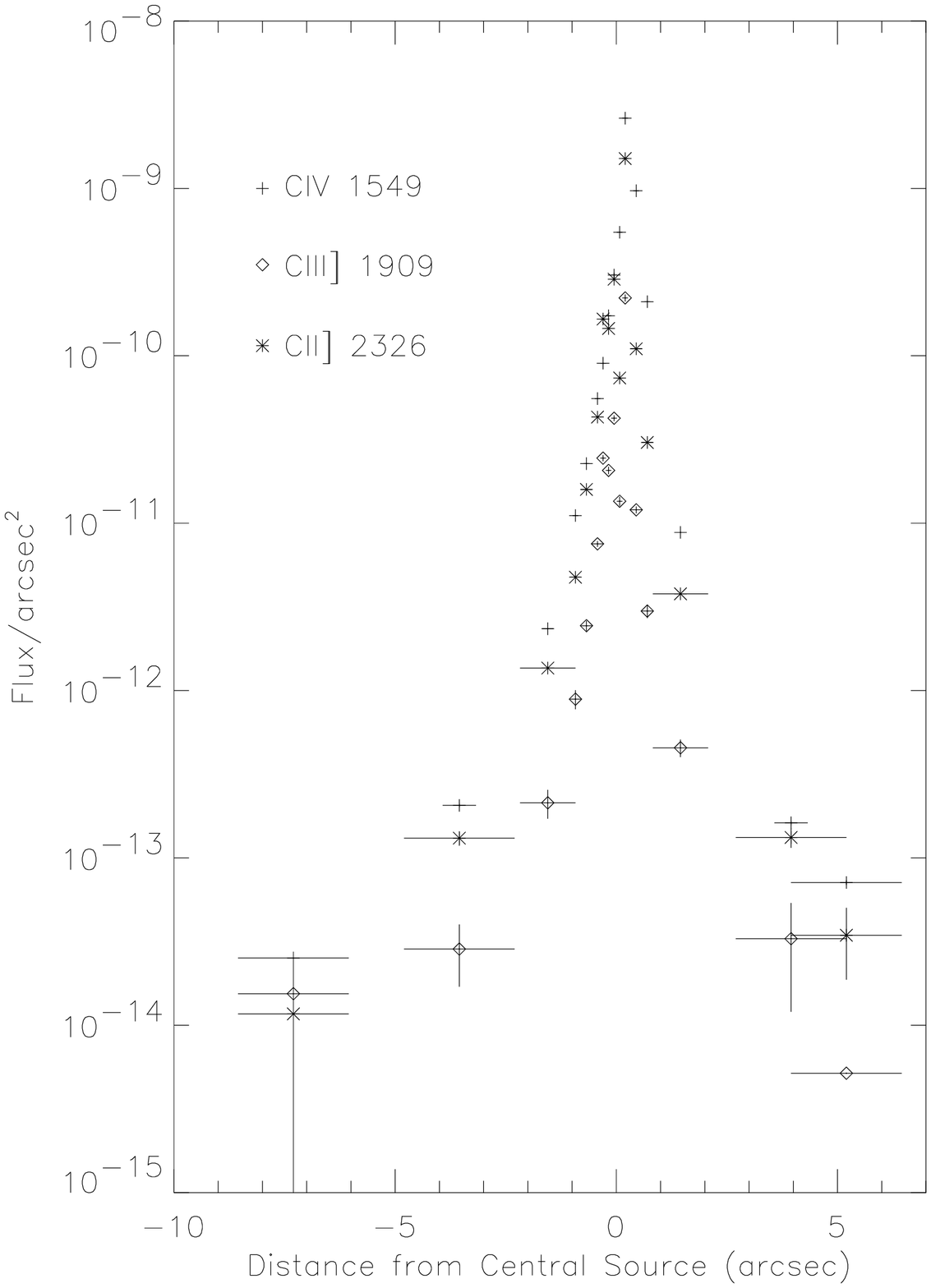}
\caption{Integrated fluxes for the C{\sc ii}] $\lambda$2326, C{\sc iii}] $\lambda$1909 and C{\sc iv} $\lambda$1549 emission lines plotted as a function of radial distance from the slit centre.}
\label{fluxes}
\end{center}
\end{figure}

The resulting spectra from both gratings, although particularly G140L, showed that a number of emission lines remained visible in the outer parts of the nebula -- we show the integrated fluxes of lines C{\sc ii}] $\lambda$2326, C{\sc iii}] $\lambda$1909 and C{\sc iv} $\lambda$1549 in Fig.~\ref{fluxes}. As we would expect, the fluxes decreased from the centre of the slit outwards to the edge of the nebula. With a spatial scale of 0.025 arcsec/pixel it is clear that the nebula maintains a significant flux level to a diameter of at least 15 arcsec and possibly as much as 20--25 arcsec, although improved sensitivity is required to confirm this. This may seem surprising when compared with the images presented earlier in this paper which span about 2 arcsec at most. However we draw attention to work by Bang et al. (1992) who found evidence for a $\sim24$ arcsec nebula from long-slit high-resolution spectroscopy of the H$\alpha$, He {\sc ii} $\lambda$4686 and [O {\sc iii}] $\lambda$5007 emission lines. The presence of this extended emission was confirmed by Corradi et al. (1999) who obtained [N {\sc ii}] $\lambda$6583 spectra and images of V1016 Cyg; in this case the nebula appeared extended in the same direction as the [O {\sc iii}] $\lambda\lambda$4959,5007 image above. This is the first time that the ultraviolet counterpart to the extended nebula has been discovered, although with only one STIS observation at a single slit position we are unable to investigate the detailed morphology of this extended ultraviolet emission.

\begin{figure}
\begin{center}
\epsscale{1}
\plotone{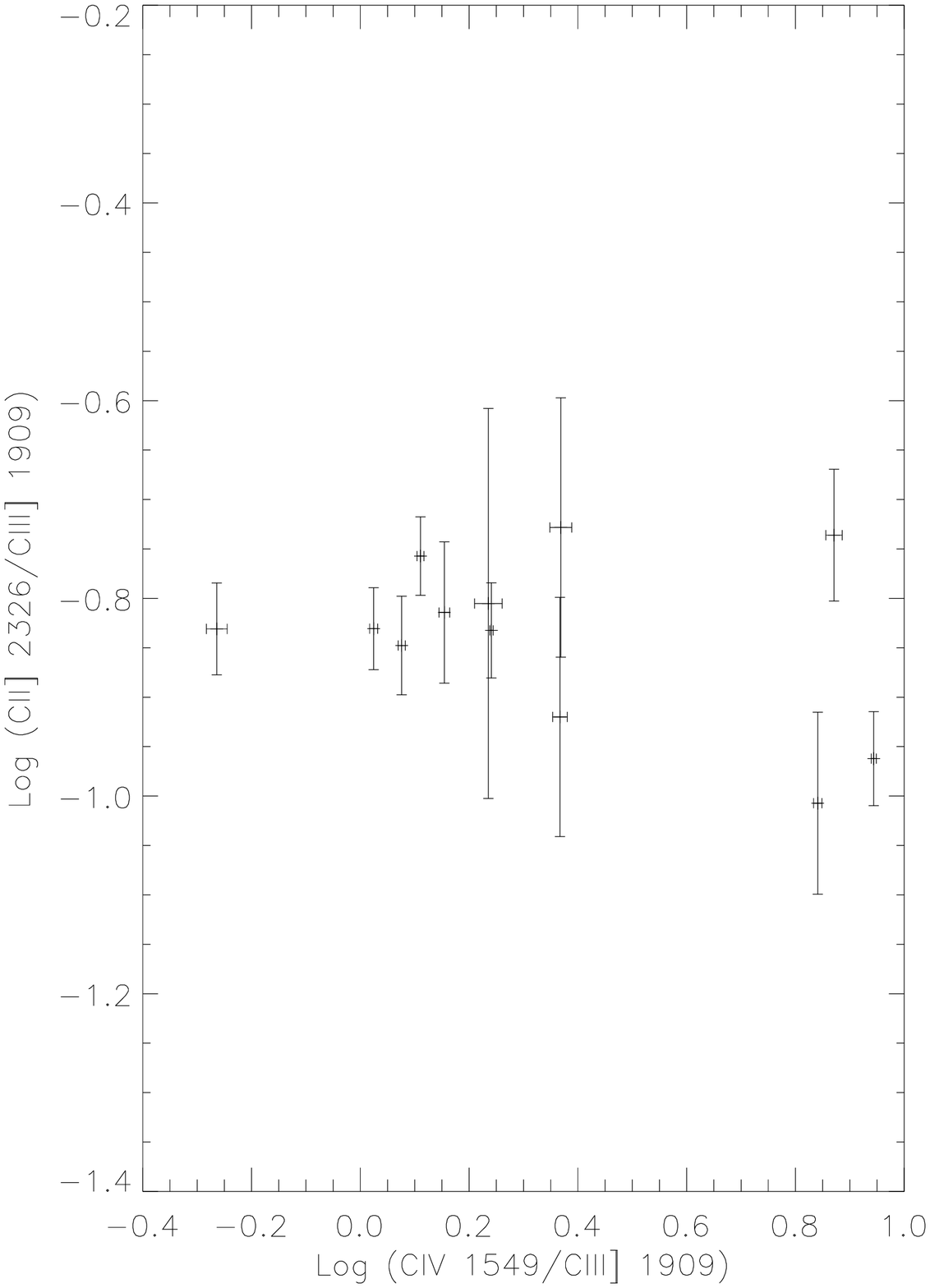}
\caption{Ratios C{\sc ii}] $\lambda$2326 / C{\sc iii}] $\lambda$1909 and C{\sc iv} $\lambda$1549 / C{\sc iii}] $\lambda$1909 plotted as for diagnostic diagram of Allen et al. (1998). With the exception of points to the edge of the nebula (for which the error-bars are extremely large) all points appear to lie in the region of photoionisation from their models, although without totally ruling out the possibility of shocks.}
\label{diagnostics}
\end{center}
\end{figure}

Finally we have plotted in Fig.~\ref{diagnostics} the C{\sc ii}] $\lambda$2326 / C{\sc iii}] $\lambda$1909 vs. C{\sc iv} $\lambda$1549 / C{\sc iii}] $\lambda$1909 ratios as for the diagnostic diagram of Allen et al. (1998, although we note that this model has been developed for active galaxies). Points from the outermost parts of the nebula have not been included due to the extremely high uncertainties in these regions. All other points lie in a region of the diagram which is suggestive of photoionisation, although the possibility of shocks is not actually ruled out. Perhaps surprisingly, there is no relationship between position in this diagram and position in the nebula, although it is possible that this will change with improved sensitivity data. These results are preliminary and more detailed modelling will take place, particularly if we obtain data with higher signal-to-noise at both optical and ultraviolet wavelengths.

\subsection{Comparison with HM Sge}

HM Sge and V1016 Cyg share a number of common properties which has led frequently to their simultaneous study (including e.g. Corradi et al. 1999, Schmid \& Schild 1990). Both systems are D-type symbiotics with a Mira as the cool component; furthermore they are both symbiotic novae which have experienced an outburst within the last forty years. They are both relatively bright radio sources and the published estimates of their orbital periods are relatively long compared with other symbiotic systems. Comparison of our results above with Eyres et al. (2001) suggests that coincidentally both central binary systems had similar separations and position angles at the time of our observations.

Unfortunately such comparisons are insufficient if we wish to study the characterstics of the sources. Despite such similar parameters HM Sge is considerably brighter and appears to have much more structure within the nebula. Furthermore there is evidence for non-thermal radio emission from extended regions whereas there is no such emission in V1016 Cyg observed to date. In addition, our preliminary ultraviolet diagnostics suggest that photoionisation is more significant than shocked emission in the V1016 Cyg nebula. The $\sim10$-year gap in time between the oubursts of these two sources may explain some of these differences. Current distance estimates to V1016 Cyg (2--10 kpc) suggest that it is considerably further away than HM Sge ($\sim$ 1250 pc, Richards et al. 1999) and this may explain some of the differences; however if, as we suggest above, the distance is actually less than those published then it would be interesting to determine more about how and why these two sources differ, rather than focussing soley on their similarities. 

\section{Conclusions}
We have presented the first sub-arcsecond optical images of the symbiotic star V1016 Cyg and have shown that although the source was approximately circular in most images, the [O {\sc iii}] $\lambda\lambda$4959,5007 emission appears to trace a bipolar outflow with axis aligned approximately NE-SW. These images have allowed us to determine the positions of the two stellar components of the system and thus a separation of 84 AU (assuming a distance of 2 kpc) and position angle of 143.5$^{\circ}$; the subsequent orbital period calculated for such a projected separation is extremely long ($\sim 544$ years) and may require revision of the distance estimate and/or a highly eccentric orbit. We have added a note of caution here, however, as we need to conduct further work to totally eliminate any possible instrumental effects. Furthermore we have presented a 23 GHz radio image which we have used in calculation of firstly the electron density and temperatures for different path lengths through the nebula and secondly the extinction at different locations within the nebula; the brightness temperature near the centre of the radio image was $5424\pm113$ K. Finally we have determined accurately the position of the slit in archival STIS spectra thus enabling us to show that there is extended emission in the ultraviolet in a $\sim15-25$ arcsec (diameter) nebula, as observed in optical data previously and which appears to be photoionised from preliminary analysis. Further STIS spectrosocopy, particularly if used to `map' the nebula, is strongly encouraged in this and other extended nebulae of symbiotic sources.

\section*{Acknowledgements}
Many thanks are due to Dr. S.J. Smartt of the UK {\sl HST} Support Unit for his assistance with extracting the one-dimensional STIS spectra. The VLA is operated by the National Radio Astronomy Observatory, a facility of the National Science Foundation operated under cooperative agreement by Associated Universities, Inc. The Hubble Space Telescope (HST) is a joint project between the US National Aeronautics and Space Administration (NASA) and the European Space Agency (ESA). CB is supported by a PPARC grant, as was SPSE in the initial phases of this work.

\end{document}